%Paper: astro-ph/9305025
%From: fan@cass153.UCSD.EDU (Xiaoming Fan)
%Date: Wed, 19 May 93 15:30:23 PDT

\magnification=1000  % character size
\centerline{}
\vskip 1.19truein
\centerline{\bf LICK SLIT SPECTRA OF THIRTY-EIGHT}
\smallskip
\centerline{\bf OBJECTIVE PRISM QSO CANDIDATES}
\smallskip
\centerline{{\bf AND LOW METALLICITY HALO STARS}
\footnote{\raise4pt\hbox{1}}
{\rm Based on observations obtained at
Lick Observatory, University of California.} } % closes centerline
\bigskip \bigskip\bigskip
\centerline{DAVID TYTLER,
\footnote{\raise4pt\hbox{2}}
   {Dept. Physics, University of California, San Diego
(TYTLER@cass155.ucsd.edu).}
{\raise4pt\hbox{,}}
\footnote{\raise4pt\hbox{3}}
 {Center for Astrophysics and Space Sciences, 0111, University of California,
 San Diego, La Jolla, CA 92093-0111.}
  XIAO-MING FAN,{\raise4pt\hbox{3,}}
\footnote{\raise4pt\hbox{4}}
{Department of Astronomy, Columbia University, New York.}
} % closes centerline
\medskip
\centerline{VESA T. JUNKKARINEN,{\raise4pt\hbox{3}}
            and ROSS D. COHEN{\raise4pt\hbox{3}} }
\bigskip\bigskip
%\centerline{Submitted to the Astronomical Journal}
\medskip
%\centerline{Received: \hbox to 1truein{\hrulefill};  Accepted: \hbox to
%1truein{\hrulefill}}
\medskip
\centerline{\bf To appear in the Astronomical Journal, August, 1993}
\vfill\eject

\centerline{\bf ABSTRACT}
\bigskip\bigskip
We present Lick Observatory slit spectra of 38 objects which were claimed to
have pronounced ultraviolet excess and emission lines. Zhan \& Chen selected
these
objects by eye from a UK Schmidt telescope IIIaJ objective prism plate of a
field at $\rm 0^h$ $0.0^{\circ }$ ($l \simeq 98^{\circ }$,
$b \simeq -60^{\circ }$). We concentrated on $m_J \simeq 18$ --19 objects
which Zhan
\& Chen thought were most likely to be QSOs at redshift $z_{em} \geq 2.8$.

Most of our spectra have FWHM spectral resolutions of about 4~\AA , and
relatively high S/N of about 10 -- 50, although some have FWHM
$\simeq 15$~\AA ~or lower S/N.
We find eleven QSOs, four galaxies at $z \simeq 0.1$, twenty-two stars and
one unidentified object with a low S/N spectrum.

The ZC lists are found to contain many QSOs at low $z$ but few at high $z$, as
would be expected.
Of eleven objects which ZC suggested were QSOs with $z_{prism} \leq 2.8$,
eight (73\%) are QSOs. But only three of twenty-five candidates with
$z_{prism} \geq 2.8$ are QSOs, and only two (8\%) of these are at $z \geq 2.8$.
Unfortunately the ZC prism redshifts are often incorrect:
only five of the eleven QSOs are at redshifts similar to $z_{prism}$.

Six of the QSOs show absorption systems, including Q0000+027A with a relatively
strong associated C~IV absorption system, and Q0008+008 (V$\simeq 18.9$) with
a damped Ly$\alpha $ system with an H~I column density of $10^{21}$ cm$^{-2}$.

The stars include a wide variety of spectral types. There is one new DA4 white
dwarf at 170~pc, one sdB at 14~kpc, and three M stars. The rest are of types
F, G and K.
We have measured the equivalent widths of the Ca~II~K line, the G-band and
the Balmer lines in ten stars with the best spectra, and we derive
metallicities.
Seven of them are in the range $-2.5 \leq$~[Fe/H]~$\leq -1.7$, while the
others are less metal poor. If the stars are dwarfs, then they are at distances
of 1 to 7~kpc, but if they are giants, typical distances will be about 10~kpc.
\bigskip\bigskip
\noindent {\it Subject Headings:}
quasars: general -- galaxies: distances and redshifts
-- stars: fundamental parameters -- white dwarfs -- surveys
\vfill\eject

\centerline{1. INTRODUCTION}
\bigskip
Zhan \& Chen (1987a,b, 1989a,b, hereafter ZC1, ZC2, ZC3 and ZC4) presented
lists of several hundred QSO candidates, which they selected by eye from
a single UK Schmidt telescope IIIaJ objective prism plate. The candidates were
chosen because they had emission lines and UV excess in the range
3200--5400~\AA , and each was assigned a reliability index, $Q = Q_1+Q_2$,
where $Q_1$ was 1, 2 or 3 for increasing strength of emission
lines, and $Q_2$ was similarly valued for increasing strength of UV excess.

We have obtained slit spectra of 38 of these QSO candidates with the Lick
Observatory 3~m telescope, on three separate occasions, during unrelated
projects. Two of the objects have $Q=6$, thirty-four have $Q=5$ and the
remaining two, with $Q=4$, are the least likely to be QSOs. Since $Q \geq 4$
for all
thirty-eight objects, all should have both emission lines and UV excess.

Table 1 is a journal of our observations, with the instrumental setup
(\S 2 below), wavelength range and integration time. We also give the
reference to the Zhan \& Chen
paper which contains the object coordinates, magnitude and finding chart.
Note that these charts have East to  the right, and that
the English translation of ZC3 lacks charts, while for the other papers the
charts are often better in the translation.

\bigskip
\centerline{2. OBSERVATIONS}
\bigskip
Our spectra were taken in support of three different observing
programs, at three different times, and with three different instrumental
setups, although in all cases we used a Cassegrain spectrograph with the
Shane 3-m telescope at Lick Observatory.
\bigskip
\centerline{2.1 {\sl Setup A: Ten z $\simeq 3$ Objects}}
\bigskip
These targets were selected as bright QSO candidates with ${z_{prism}} \simeq
3$.
They were selected and observed by VTJ and RDC.

The UV Schmidt camera was used on the Cassegrain spectrograph (Miller \& Stone
1987) with a 300 g/mm grating blazed at 4230~\AA ~in first order. A thinned TI
800$\times$800 CCD with 15 $\mu $ pixels, and 7 e$^-$ readout noise was used,
giving
3.9~\AA ~per pixel across 3100~\AA . All observations were made on October 4,
1988, when the sky was clear and the seeing was about 1 arcsecond. For the ten
program objects we used a
wide 2.88 arcsecond slit, giving a FWHM resolution of 3.5 pixels,
which is 14~\AA , but for the flux standard star we used a
7.9 arcsecond slit.
The slit was not rotated to the parallactic angle, but it
was rotated to $PA=248^{\circ }$ to simultaneously record 0003+011A \& B, and
to
$PA=92^{\circ }$ for 0011$-$002A \& B. Hour angles ranged from 2 hours 30
minutes East to 3 hours 10 min West. All exposures were 600 seconds, and the
spectra were reduced in the usual way.
\bigskip
\centerline{2.2 {\sl Setup B: Eleven Intermediate z Objects with Close
Neighbours}}
\bigskip
These eleven QSO candidates were observed on August 24 or 27, 1990 by DT and
FXM
to search for Mg~II absorption systems which might show large scale
($\simeq 100~h^{-1}$~Mpc) correlations in three dimensions. This program was
motivated by the finding of Tytler et al. (1987) that a few QSOs each had more
Mg~II systems than were expected if they were all intervening,
a two sigma result which has since been refuted by much larger samples which
do not show any sign of such correlations (Sargent et al. 1988;
Steidel \& Sargent 1992, Tytler, Sandoval \& Fan 1993 \S 2.4). The QSO
candidates which we observed were chosen because they had one or more
neighbours within
about $1^{\circ }$. Here we present our observations of only the ZC
QSO candidates, two of which were also observed by VTJ and RDC with setup A.
Other QSOs observed in this program will be discussed elsewhere.

Spectra were obtained with the Cassegrain spectrograph
using a 600 g/mm grism, blazed at 4840~\AA ~in first order (Miller and Stone
1987). We used a thinned TI
800$\times$800 CCD with 15 $\mu $ pixels, and 7~e$^-$ readout noise, giving
3.43~\AA ~per pixel from 4312 to 7059~\AA . The wavelength range was chosen to
maximize the chance of detecting redshifted Mg~II absorption line systems, and
it unfortunately misses blue wavelengths which are most useful for stellar
spectral classification.
A 2.09 arcsec slit was used giving a FWHM resolution of 2.5 pixels,
which is 8.6~\AA . The slit was not rotated to the parallactic angle, but the
spectra were reduced in the usual way.
\bigskip
\centerline{2.3 {\sl Setup C: Nineteen High z Objects}}
\bigskip
Nineteen QSOs were observed by DT in November 1992 with the superb new Kast
double spectrograph, which records blue and red spectra simultaneously.
We used a dichroic with a nominal wavelength of 5500~\AA .
Light with wavelengths to the blue of this were dispersed with a 600 g/mm
grism blazed at 4310~\AA , while the red light was dispersed by a 600 g/mm
reflection grating blazed at 7500~\AA . A thinned Reticon 1200$\times$400 CCD
was used in each of the blue and red
cameras. In the blue we recorded from 3320 to 5485~\AA ~with 1.81~\AA
{}~per pixel, and a two pixel FWHM of 3.6~\AA , and in the red from 5530 to
8270~\AA ~with 2.34~\AA ~per pixel and a two pixel FWHM
of 4.7~\AA . A 1.5 arcsec slit was used, and was rotated to the appropriate
parallactic angle. The sky was clear, the seeing about 1.5 arcseconds, and
the spectra were reduced in the usual manner.
\bigskip
\centerline{3. RESULTS}
\bigskip
We first discuss various problems with the spectra, then slit magnitude and
color estimates.
\bigskip
\centerline{3.1 {\sl The Spectra}}
\bigskip
The spectra shown in Figure 1 are grouped by setup (A, then B, then C)
to make spectral features easier to identify. The flux is $f_\nu $, in units
of micro-Jansky, and all wavelengths given in this paper are vacuum, but they
are not heliocentric.
Wavelength scales should be accurate to about one pixel or better.
A one sigma error trace is shown beneath the spectra from setups B and C. Peaks
in the error correspond to sky emission lines. Note that poorly subtracted sky
emission lines can appear as emission and/or absorption in the spectra.
In setup B there is frequently a bogus absorption at the extreme blue end of
the spectrum (4311~\AA ), a bogus emission feature near 4325~\AA, and a second
bogus absorption near 4370~\AA . The former two arise from poor flux
calibration, while the third is bad sky subtraction.

For setups A we did not attempt to correct for atmospheric absorption,
hence the B band (6867~\AA ) is visible. For setups B and C we did use early
type star spectra to attempt to remove the B band, the A band (7600~\AA ), and
OH absorption at 7160--7340~\AA ~but with varying success.

The spectra have been corrected for atmospheric extinction, but not for
interstellar extinction ($b \simeq -60^{\circ }$).
Table 2 is a summary of our results.
\bigskip
\centerline{3.2 {\sl Slit Magnitudes}}
\bigskip
Magnitudes listed by ZC were obtained from image sizes on a direct Schmidt
plate, using the King et al. (1981) calibration of the dependence of $B_J$
magnitude on image size. ZC5 noted that these magnitudes may
be too bright by 0.5 -- 1.0 mag. because
they found most objects at $\simeq 18.5$, a whole magnitude brighter than the
peak of the otherwise similar survey by Savage et al. (1984). However
H$\rm \ddot o$rtnagl, Kimeswenger \& Weinberger (1992) have shown that
King et al. measured larger image diameters at a given $m_j$, which suggests
that the ZC magnitudes may actually be too faint, rather than too bright, for
$m_j \geq 18$. We can not determine which is correct because we do not know
how ZC measured image sizes.

To try to reduce this uncertainty, we have estimated magnitudes
from our slit spectra. These magnitudes are highly uncertain because we used
narrow slits. A broad band flux $F$ is defined as
$$ F = {{\int^{\infty }_{\lambda _{min}} T(\lambda )f(\lambda )d\lambda }\over
{\int^{\infty }_{\lambda _{min}} T(\lambda ) d\lambda }}, \eqno(1)$$
where $T (\lambda )$ is the band transmission (Kitchin, C.R. 1984),
$f( \lambda )$ is the flux per unit wavelength, and
$\lambda _{min}$ is ideally $ -\infty $, but in practice was the minimum
wavelength of the spectrum. For setup C we added an estimate of the flux
in the dichroic filter gap, which was only 20~\AA . We obtained magnitudes from
$U = -2.5log_{10}F_U + K_U$, and similarly for B and V, where the constants
$K_U$, $K_B$ and $K_V$ were obtained from the standard star spectra.

Our slit magnitudes are listed in Table 3. They have automatically been
corrected for atmospheric extinction by the usual flux calibration process,
which converts from
recorded photoelectrons to flux above the atmosphere, as a function of
wavelength. Galactic reddening E(B$-$V) values from Burstein \& Heiles
(1982) are 0.01, 0.02, or 0.03 for the targets.
The colors (B$-$V)$_0$ listed in Table 3 have been corrected
for these reddening values, but we have not corrected the individual magnitudes
because their zero point errors are much larger than the corrections of
A$\rm _V \simeq 0.03 - 0.09$.

We have checked our magnitudes in five ways. First, we calculated a
magnitude from each of our thirteen standard star spectra.
We obtained standard deviations of $s=0.17$ magnitudes
for the V magnitudes and $s=0.21$ for the B magnitudes, which we regard as
lower bounds on the external errors of our other magnitude measurements.

Second, two objects were observed twice, and in both cases the
magnitudes were 0.7 -- 1.0 fainter in setup A, because a
wider slit was used for the flux standard than for the program objects,
although the (B$-$V) colors differ by much less (0.02 and 0.14).

Third, when we compare our slit (B$-$V) colors with those
estimated from the strength of the Balmer lines in \S 3.4 below, we find
excellent agreement
which suggests that our slit (B$-$V) colors from setup C (the only ones to have
the silt aligned to the parallactic angle) have a 1$\sigma $ error of under
0.04 mag.

Fourth, in Figure 2 we show the difference
between our slit B magnitudes and the ZC image size B$\rm _J$ magnitudes. Ours
are on average 1.4 magnitudes fainter. ZC5 noted that their magnitudes were
probably too bright, but they guessed by only 0.5 -- 1.0 magnitudes, which
suggests that some of our magnitudes may be too faint. We do not see any
systematic differences between our three setups.

Fifth, four of the QSOs (0004-005B, 0006+020B, 0006+025 and 0010-002B) have
been found independently by Foltz et al. (1989).
Their $B_J$ magnitudes, which we list in \S 6,
are brighter than ours by 0.3, 0.54, 1.19 and 1.57 magnitudes respectively,
where the first three are from setup B, and the last one is from setup A.

These tests suggest that our colors, but not necessarily the  individual
magnitudes, from setup C are good. Both the magnitudes and
colors from setups A and B setups are also suspect because
the slit was narrow (1.5 to 2.88 arcsec), it was not rotated to the
parallactic angle, and the TV camera guides on the red, so
we expect that the B and especially the U magnitudes will be systematically
too faint, and the V magnitudes should be the least bad. In addition there are
several reasons why we expect our magnitudes for the program objects to be
systematically too faint. The standard stars were
observed for much shorter times than the faint targets, they should be
better centered on the slit, and better focused, and for setup A a wider slit
was used for the standard stars than for the program objects.
\bigskip
\centerline{4. STARS}
\bigskip
Papers by Gunn \& Stryker (1983) and
Jacoby, Hunter \& Christian (1984) were consulted to obtain rough stellar
classifications.
Berg et al. (1992) present a simple classification scheme which they used on
their 6~\AA ~ FWHM optical spectra of QSO candidates.
Beers et al. (1992a) present digital spectra covering 3700 -- 4500~\AA ~at
0.7 to 1.2~\AA ~FWHM for various hot halo stars (A, DA, sdO, sdB, Horizontal
Branch = HB), together with classification criteria, while
Beers, Preston \& Shectman (1992b, hereafter BPS2) present spectra of cooler
halo F and G stars of various metallicities.

Greenstein (1980) discusses the difficulty of distinguishing white
dwarfs (WDs) from hot halo (subdwarfs sdO or sdB, hot HB) stars.
The separation is hardest for the hottest stars because the physical
differences
in temperature and gravity are also small, so that DAwk (weak Balmer lines) and
sharp lined DAs can be confused with sdBs. But the distinction is easy below
12,000~K because
the WDs then have stronger Balmer lines. Greenstein (1980) shows that any star
with W(H$\gamma $) $\geq 15$~\AA ~must be a DA, but as W(H$\gamma $) drops
from 15 to 5~\AA , one has either DA stars of increasing T, or stars from
the sequence HBA (HB type A), HBB, sdB and sdO, which is one of both increasing
T and
gravity. For the hot stars we searched for but did not find any
He~I 4026, 4388, and 4471, or He~II 4686.

Our spectral classifications are given in Table 2. For the stars we list the
spectral type implied by the spectral features, and then, in parentheses, that
which would correspond to the (B$-$V) if the star had solar metallicity.
Most of the stars actually have significantly lower metallicities, so their
(B$-$V) colors are from 0.1 to 0.3 magnitudes bluer (e.g.
Beers et al. 1990; hereafter BPSK) than those of
solar abundance stars with the same $\rm M_V$.
This deblanketing effect accounts for why the spectra type deduced from the
colors are hotter than those from the spectral features.
Notes on the classification of individual objects are given in \S 6 below.
\bigskip
\centerline{4.1 {\sl Stellar Metal Abundances}}
\bigskip
BPSK discussed a method of determining stellar
metallicities which is a refinement of that first presented by Preston (1959).
The equivalent width of the Ca~II~K line is used as the metallicity estimator,
with weaker lines indicating lower metallicities. Since the strength of Ca~II~K
also drops with the increasing stellar temperature, a temperature
indicator, such as the Balmer line equivalent widths or a color index, is
needed.

BPSK note that this method has several advantages: Ca~II~K is relatively
independent of gravity,
so one does not need to know whether the star is a dwarf or giant, [Ca/Fe] is
remarkably free of scatter at a given [Fe/H], and the resulting abundances
obtained from the Beers et al. 1~\AA\ resolution spectra
have an impressively small scatter of $\delta $[Fe/H] $\simeq 0.15$ dex for
$0.33 \leq B-V \leq 0.85$ (F0 -- K1)
and $-4.5 \leq$ [Fe/H] $\leq -1.0$. Even for cooler stars with
$0.85 \leq B-V \leq 1.1$ (K1 -- K5) the scatter is only 0.19 dex.

We have measured equivalent widths of lines in our spectra using the
prescription of BPSK
(their Table 1 and \S 4c). We used only the 18~\AA ~bandpass for Ca~II~K and a
12~\AA ~bandpass for the Balmer lines. The index KP$^\prime$ is the Ca~II~K
equivalent width
measured in the 18~\AA ~interval, corrected for interstellar Ca~II~K absorption
by subtracting 0.22~\AA ~($\rm =0.192~\AA ~/sin 60{^\circ }$ from BPSK,
but see also Bowen 1991). The Balmer line index
$$ HP = -0.120 +0.5 H\delta +0.555H\gamma , \eqno(2) $$
is the average of the measured equivalent width of H$\delta $, and an estimate
of that width based on the measured equivalent width of H$\gamma $.
The G-band (CH) index GP is the equivalent width in the 15~\AA ~band around
4300~\AA .

We report values for these indices in Table 3. The Balmer line equivalent
widths can be used to estimate stellar temperatures and colors. BPS2
(their Fig. 7, and eqn. 1) found that the relationship
$$ (B-V)_0 = 0.962 - 0.292HP +0.036HP^2 \eqno(3)$$
applies for $0.35 \leq (B-V)_0 \leq 0.9$, (F2 -- K2), with most sensitivity for
the hottest stars. These estimates of $(B-V)_0$ can be compared with those
from the slit photometry. The slit minus HP color differences, listed as
$\Delta $ in Table 3, are unexpectedly
small, with a mean of $+0.008$ and a standard deviation of only 0.04, which is
only slightly worse than the prediction error of 0.$^m$03 quoted by BPS2 for
their data.
We had expected that our results would show much more scatter because we do
not have proper photometric colors and we used a lower spectral resolution of
3.6~\AA\ versus 1~\AA , but apparently these are not serious deficiencies.

Metallicities can now be measured from Fig. 4 of BPSK, which shows KP$^\prime$
as a function of (B$-$V)$_0$ for various abundances. In Table 4 we list
abundances appropriate for giants, subgiants, and dwarfs.
If the stars are dwarfs then their abundances
will be lower by up to 0.2 dex.

The metallicity errors that we list come entirely from the two different
estimates of (B$-$V). The actual random errors are probably about 0.3 dex and
are dominated by the uncertainty in the Ca~II~K line equivalent widths.
The main systematic error is probably an underestimate of metallicities
of those stars with strong Ca~II~K. When Ca~II~K is very strong
it spills outside the band pass defined by BPSK, and will will have
measured a systematically smaller KP$^\prime$, and hence
a lower metallicity, than BPSK because our spectra are of lower resolution.
We have not attempted to correct this bias, which could be done by
calibrating the KP$^\prime$ versus (B$-$V) relationship for spectra with our
resolution.

Two objects lie just outside the color range
considered by BPSK. For 0009$-$003 we use the linear extrapolation
(``patch'') for [Fe/H]$\geq -1.0$ discussed in \S 5 and presented in
Table 4 of BPS2, while for 2358+004 we extrapolate by eye to obtain
[Fe/H] $\simeq -0.5$.

The metallicities listed in Table 4
range from $-$0.5 to $-$2.6, but 50\% of the stars have
$-1.9 \leq$[Fe/H]$\leq -1.7$
which is reasonable for faint (halo) stars selected to have UV excess.
\bigskip
\centerline{4.2 {\sl Stellar G-Band Strength}}
\bigskip
We have measured the strength of the G-band CH feature in the spectra of the
ten stars for which we have determined abundances. Beers, Preston \& Shectman
(1985, hereafter BPS1) found a range of equivalent width for the GP index
which was similar to that found by others for globular cluster and halo stars,
with weak G-band stars most often on the red HB to asymptotic giant branch, and
strong G-band stars amongst the subgiants (Bond 1980).

BPS1 plot the distribution of G-band strength as a function of color for 105 of
their stars with [Fe/H] $\leq -2.0$. Relative to this sample, many of our
ten stars have unusually large GP. For example, 40\% of our stars, but only
1\% of theirs have GP $\geq 4.0$. This is partly because some of our stars have
larger abundances, and partly because we lack the bluest of stars with
(B$-$V)$ \simeq 0.4$, and we have an excess of redder stars (which have
stronger
GP), but it does still seem that we might have a few stars including
0003+012B and 2357+009 with anomalously strong CH. Our approximate colors
suggest that both of these stars
are too red to be subgiants, but not by much: 0003+012B has (B$-$V)
$\simeq 0.58$, and 2357+009 has (B$-$V)$\simeq 0.63$.
\bigskip
\centerline{4.3 {\sl Stellar Distances}}
\bigskip
If we knew whether our stars were giants or dwarfs we could obtain rough
estimates of their photometric distances, but our spectra and slit colors are
not decisive.

The relative proportion of giants (absolute magnitude brighter then the main
sequence turnoff of M$\rm ^v = 4.5$) to dwarfs depends on three factors: the
ultraviolet excess
bias in favor of low abundance, the ratio of giants to dwarfs in the disk and
halo, and the apparent magnitudes of the stars.

Seven of our ten stars with the lowest abundances
($-2.5 \leq [Fe/H] \leq -1.7$) are likely to be in
the halo because their abundances are too low for disk stars, and there are few
halo stars currently passing through the disk. BPS1 argued that all of their
low
metallicity stars were halo objects, and they used the Bahcall-Soneira
model of the Galaxy to estimate that about 90\% of these
were giants, favored because they sample a larger volume.

Our stars are much fainter than theirs, $18.1 \leq V \leq 19.7$, compared with
$12.0 \leq V \leq 15.0$. The relative size of the volume sampled for dwarfs and
giants remains unchanged with changing magnitude limit, but our sample will
include a larger proportion of
dwarfs because the density of dwarfs in the inner halo drops slower than
that of giants in the outer halo.

Our estimates of the distances to our stars are given in Table 4. We use
the absolute magnitudes specified in Table III of BPSK for dwarfs and giants
separately. These distances range from 1.4~kpc to 7~kpc if the stars are
dwarfs,
and from 7 to 62~kpc if they are giants, but in either case only three stars
could be beyond 15~kpc. Of these three, two have distances of 43~kpc and
62~kpc if they are giants. Stars are sufficiently rare at these distances
that these two objects are much more likely to be dwarfs.
Many of the stars are near the main
sequence turnoff so their absolute magnitudes and hence their distances are
less sensitive to whether they are dwarfs or giants, provided they are not
horizontal branch stars.
\bigskip
\centerline{5. QSO EMISSION AND ABSORPTION LINES}
\bigskip
QSOs were identified by the presence of at least two emission lines, one of
which might appear in the ZC spectra rather than ours because of the
differences in wavelength coverage.

We list the wavelengths and observed frame equivalent widths for the QSO and
galaxy emission lines in Table 5.
Redshifts were weighted by the line equivalent widths.
Absorption lines in the spectra of the QSOs are listed in Table 6.
Individual cases are discussed in the next section.
\bigskip
\centerline{6. NOTES ON INDIVIDUAL OBJECTS}
\bigskip\noindent
{\bf 0000+025A: QSO. -- } The spectrum clearly shows that this is a QSO
with $z_{em}$=1.6843. Absorption
lines at $\lambda$4343, $\lambda$4374 and $\lambda$5897\AA\
are probably caused by bad subtraction of strong sky emission lines.
A possible broad absorption feature at $\lambda$6379 is unidentified.
\bigskip\noindent {\bf 0000+027A: QSO. --} ZC1 identified emission lines at
4203 and 5179~\AA ~as C~IV and C~III] respectively. They are actually Ly$\alpha
$
and C~IV, giving a higher redshift. There is a strong absorption line just to
the red of the peak of the C~IV emission line which is almost certainly
associated C~IV absorption, with $z_{abs}=2.394$ and a rest frame equivalent
of 3.0~\AA ~because of its strength and position.
\bigskip\noindent {\bf 0003+011A: star -- M4. --} The spectral features suggest
M4 star, while the (B$-$V) is typical of a M2 star of solar abundance.
\bigskip\noindent {\bf 0003+011B: star -- F. --} The continuum falls of
shortward of 5000~\AA , and rises slightly into the red, indicating
$T \leq 7000$~K. Ca~II H and K are weak, and there is no break at 4000~\AA .
\bigskip\noindent {\bf 0004$-$005B: QSO. --}
This QSO was identified as Q0004-0032 with $z_{em}=1.72$ and $B_J=18.4$
by Foltz et al. (1989). Their spectrum shows four strong emission lines.
Two strong absorption lines at $\lambda\sim$5854 and
$\lambda\sim$5959 are unidentified. They might be Mg~II absorption systems.
\bigskip\noindent {\bf 0004+014: star -- G. --} Ca~II H,K and Mg~I triplet
5167,
5172.7, and 5183.6  are very strong. H$\alpha $ which is very weak, is the only
visible Balmer line. Flux drops about 50\% across 4000~\AA .
\bigskip\noindent {\bf 0005+030: QSO. --} ZC1 identified an emission line at
4000~\AA ~as Ly$\alpha$. It is actually C~III]~1909, which is confirmed by our
detection of Mg~II at $z_{em}=1.0948$ and blended Fe~II features.
\bigskip\noindent {\bf 0006+020B: QSO. --} This QSO was identified as
Q0006+0200 with $z_{em}=2.35$ and $B_J=17.9$ by Foltz et al. (1989).
ZC1 identified emission lines at
4149 and 5200~\AA ~as C~IV and C~III] respectively. They are actually
Ly$\alpha $ and C~IV, at a higher redshift of 2.3483.

The two strong absorption lines in this spectrum at 4700\AA\ and 5226\AA\
and a possible weak line at 5076\AA. The absorption feature at 5578\AA\ is
considered bogus because it is near to a strong sky line.
The strong line at 4700\AA\ and the weak line at 5076\AA\ could be
C~IV and Al~II at $z_{abs}$=2.035, but the Si~II $\lambda$1526 line which
would be expected at 4634\AA\ is not seen, and the second strong line is not
identified. Alternatively, 4700 is Si~IV$\lambda$1393, 4735\AA\ is
Si~IV$\lambda$1402,  and 5226\AA\ is a blended
C~IV doublet. But then 4700\AA\ must be a blend with another unidentified line
because it is twelve time stronger than its doublet partner, compared to a
maximum ratio of two. The Foltz et al. (1989) spectrum also shows the
4700 line. In a footnote to their table 2 they suggest possible associated
absorption, presumably because they see absorption in the
Ly$\alpha $ emission line, which is consistent with our interpretation of
the 5226 line as C~IV.
\bigskip\noindent {\bf 0006+022A: QSO. --} ZC1 identified emission lines at
3910 and 5056~\AA\ as Ly$\alpha$ and C~IV respectively. They are actually C~IV
and C~III] respectively, at $z_{em}=1.5152$. We also observe the blue wing of
the Mg~II emission line which should be centered at about 7039~\AA .
\bigskip\noindent {\bf 0006+025: QSO. --} Foltz et al. (1989) called
this QSO Q0006+0230. They saw
Ly$\alpha $ and Si~IV+O~IV] emission lines in addition to the C~IV and C~III]
which we saw, and obtained $z_{em}=2.09$, consistent with our value of
$2.0909$. Their magnitude of $B_J=18.00$
is to be preferred to our value of $B=19.19$.
\bigskip\noindent {\bf 0008+008: QSO. --} The peak of the Ly$\alpha $ emission
line appears at too high a redshift because strong absorption lines
destroy the blue side of the line peak.
The colors listed in Table 3 are both much redder than those of typical QSOs
because of the Ly$\alpha $ forest and the Lyman limit system.

A Lyman Limit edge is seen at $\lambda\sim$3750 which corresponds to
$z_{LLS}$=3.08. This system at $z_{abs}=3.079$
is confirmed by strong corresponding
Ly$\alpha$ and C~IV doublet absorption lines. In addition there is an obvious
damped
Ly$\alpha$ absorption line at $\lambda\sim$4883.5~\AA\ ($z_{abs} \simeq 3.017$)
with an observed equivalent width of about 96\AA , which corresponds to
a neutral hydrogen column density of
N(HI) $\sim$ 10$^{21}$ cm $^{-2}$. The metal line absorption system with
$z_{abs}$=3.028 is likely to be associated with this damped Ly$\alpha$
line. Other metal systems including C~IV at $z_{abs}$=2.625,
2.650, 2.895, and Mg~II at $z_{abs}$=1.194 were found.
\bigskip\noindent {\bf 0008+010: star -- G. --} This star shows H$\alpha$,
H$\beta$, H$\gamma$ and Ca~II, all of which are very weak. There is no 4000~\AA
{}~break and Ca~II is stronger than Balmer lines. The estimated metallicity
of [Fe/H] $\simeq -2.5$ is the lowest of our ten stars.
\bigskip\noindent {\bf 0010$-$002B: QSO. --} ZC3 listed a weak emission line
at 4203\AA , a strong line at 3818~\AA\ which they interpreted as Ly$\alpha$
(rest wavelength 1228~\AA) at an emission redshift of 2.11. We see weak
emission lines at 4872\AA\ and 6004\AA\ which we interpret as C~IV and C~III]
at a redshift of 2.1462. A Foltz et al. (1989) spectrum of this object,
which they call Q0010$-$0012, shows Ly$\alpha $ in
addition to the weak C~IV and C~III] lines which we saw. They obtained
$z_{em}=2.15$, which is consistent with our redshift. They gave B$_J=18.5$,
which is more reasonable than our $B=20.07$.

ZC3 also list an absorption line at 3669~\AA, which is likely to be a
strong (probably damped because it shows up in the prism spectrum)
Ly$\alpha$ line at $z_{abs}=2.017$. The Foltz et al (1989) spectrum also shows
this line.
We see a partially resolved pair of lines at 4690\AA\ and 4697.7\AA,
which we interpret as the C~IV doublet in this absorption system,
with $z_{abs} =2.030$ (see also the Foltz et al spectrum).
We also see a possible doublet near 6300\AA\ ~which
might be Mg~II if it is not simply poor sky subtraction.
\bigskip\noindent {\bf 0010+008: QSO. --} There is a possible Lyman limit
edge at about 3500 -- 3600\AA\ but the S/N is too low to see possible metal
lines in this system.
\bigskip\noindent {\bf 0010+023: galaxy. --} This emission line galaxy
($z_{em}=0.0879$) was observed in both setups A and B. We see stellar
absorption immediately surrounding the Balmer emission lines in the setup
B spectrum.
\bigskip\noindent {\bf 0011$-$002A: galaxy. --} In addition to the
emission lines listed in Table 5, the 4000~\AA\ break is also seen
at the expected wavelength of 4463~\AA .
\bigskip\noindent {\bf 0011$-$012A: star -- sdB. --} The Balmer lines are
resolved
and have large equivalent widths of W(H$\alpha$) = 12.1~\AA , W(H$\beta$) =
13.1~\AA , and W(H$\gamma$) = 3.9~\AA . The spectrum rises to
the blue, indicating T$ \geq 11,000$~K, but the slit (B$-$V) of 0.18 indicates
a much lower temperature of T$\simeq 8,000$~K.

The absorption feature at the extreme blue end of the spectrum near
4311~\AA , and the shallower broad feature at 4370~\AA ~are both bogus, the
latter lying in the region of He~I~4387. However He~I 4471~\AA , which is
normally stronger, is not seen. H$\gamma $ at 4345~\AA ~may be compromised by
these detector problems.

The W(H$\gamma $) line has a central depth of $R_c =0.33$ of the continuum
level
(measured down from the continuum, so that deeper lines have larger values), a
FWHM of 10.5~\AA , and a width at 20\% below the local continuum of
$D(0.2) = 8.4$~\AA . Both the width measurements have been corrected
for the instrumental resolution, but we have not
corrected the central depth, which should be greater than the measured value.
The corrected width and uncorrected depth place the star amongst sdO, sdB and
hot DA stars of Figure 2 of
Beers et al. (1992). The line profile lacks the extensive wings expected of a
white dwarf, but is very similar to a hot sdB (or cool
sdO) star with T=35,000~K and log~g = 5 (Fig. 3 of Greenstein 1980). Since the
line will be deeper than our measurement, the actual temperature is probably
nearer to 25,000~K, and the star would be near to the blue HB.
However these high temperatures are inconsistent with the (B$-$V),
and neither the H$\gamma $ line nor the color can be considered reliable.

We can also measure the width of the H$\beta $ line at 0.9 of the continuum
level,
which increases with both temperature and gravity (Herbig 1992). The measured
value of 36~\AA ~implies $\rm 18,000 \leq T (K) \leq 26,000$ if the star
is an sdB, since such stars have $5.0 \leq$ log g $\leq 5.7$, which is
consistent
with the H$\gamma $ profile, but not with the (B$-$V) color.

If it is an sdB (or blue HB) with an absolute magnitude of about M$_v \simeq
+2$
(Greenstein \& Sargent 1974), then it is in the halo at a distance of
about 14~kpc, well beyond the disk sdB stars for which Heber (1986) derived
a tentative scale height of 170 -- 220~pc.
\bigskip\noindent {\bf 0012$-$002: QSO?. --} The (B$-$V)=0.40 is typical of a
QSO or hot (F3) star. There is a probable weak emission line at 4882~\AA ~which
is probably the moderate strength line which ZC2 note at 5076.
ZC2 reported a second moderate line at 4028 which
is outside our wavelength range. They interpreted these lines as C~IV and
L$\alpha $ giving a poor fit to $z_{prism}=2.28$. We prefer $z_{em}=1.557$
which identifies the lines as C~III] and C~IV, but this is very uncertain.
\bigskip\noindent {\bf 0012+011: star?. --} Only the Na~I~D line is seen in
this
very low S/N spectrum, and even this is uncertain because it lies
on top of the strong sky emission line. The (B$-$V)=1.08 is too red for a
normal QSO, but resembles a cool (about K4) star, which would show strong
Na~I~D. The absence of absorption around 5180~\AA\ suggests
that the star can not be cooler than K4, unless its metallicity is low, which
is possible.
\bigskip\noindent {\bf 2350-019A: DA4 white dwarf. --} This new white dwarf
shows strong Balmer lines up to H5. Its continuum flux peaks between 4200 and
4800\AA , which implies temperatures of 10,600 to 12,100~K, corresponding to
DA5 to DA4 (Dx, where x $ \equiv 50,400/T$~K; Sion et al. 1983). The slit
(U$-$B) and (B$-$V) colors are consistent with DA3 -- DA5, and the
H$\beta $ line strength and profile correspond to DA3 to DA5.
The line equivalent widths are: W(H$\alpha$) = 95.7~\AA ,
W(H$\beta $) = 55.6~\AA , W(H$\gamma$) = 51.8~\AA ,
and W(H$\delta$) = 11.5~\AA .
Greenstein (1980) shows that the huge equivalent width of W(H$\gamma $) means
that the star must be a white dwarf, and that the temperature should be in
the range 12,000 -- 15,000~K (DA4 -- DA3). The spectrophotometry of Greenstein
(1984) suggests that this star would have (G$-$R) $\simeq -0.14$ and
M$_v \simeq 12.6$, which implies a distance of 170~pc, less than the
white dwarf disk scale height of about 220 -- 270~pc (Green, 1980)
\bigskip\noindent {\bf 2359+026: unclassified. --} The S/N is very low and no
definite features are seen. The apparent emission features at $\lambda 4159$
and $\lambda 5065.6$ are both considered bogus because both lie near CCD
defects, while the latter is near a strong sky line. Our slit colors for this
object,
(B$-$V)=0.25, (U$-$B)=$-$0.69, suggest that it is hot hotter in (U$-$B) than
any of the halo stars shown in Figure 6 of BPS2, and are consistent with a
DA5 -- DA6 WD, although the colors are about 0.1 magnitudes redder in
(B$-$V) or bluer in (U$-$B) than most WDs. It could also be a BL Lac object
since the colors are very near the mean for QSOs.
\bigskip
\centerline{7. DISCUSSION}
\bigskip
Zhan \& Chen have noted (ZC5) that the fraction of their QSO candidates with
$z_{em} \simeq 3$ was too large, and they commented that this was partly due to
their interest in finding such high $z$ objects.

We have found that only two of twenty-five
candidates with $z_{prism} \geq 2.8$ are high redshift QSOs (8\%). Their
success rate is much better for lower $z_{prism}$, with eight QSOs out of
eleven
candidates (73\%). Unfortunately the ZC prism redshifts are often incorrect
because of incorrect emission line identifications. Only five of the eleven
confirmed QSOs are at redshifts within
0.5 of the $z_{prism}$ values, although for these five the $z_{prism}$ were
very
good, differing from $z_{slit}$ by only 0.03 -- 0.06.

We wish to thank George Preston for advice on the stellar spectra,
the Lick night assistant Jim Burroughs,
for help with the observations at Lick Observatory.
This work was supported in part by NASA grants NAGW$-$2119 (FXM \& DT),
GO$-$3801.01$-$91A from the Space Telescope Science Institution, which is
operated by AURA Inc. under NAS5$-$26555 (FXM \& DT), and contract NAS
5$-$29293 (VTJ, RDC).

\vfill\eject
\bigskip
\centerline{REFERENCES}
\bigskip
\def\jref#1#2#3#4{{\par\noindent\hangindent=3em\hangafter=1 #1, {#2},
{#3}, #4\par}}
\def\ref#1{{\par\noindent \hangindent=3em\hangafter=1 #1\par}}
\jref{Beers, T.C. 1990}{AJ}{99}{323}
\jref{Beers, T.C., Preston, G.W., Shectman, S.A. \& Kage, J.A. 1990}{AJ}{100}
{849 (BPSK)}
\jref{Beers, T.C., Preston, G.W., \& Shectman, S.A. 1985}{AJ}{90}{2089 (BPS1)}
\jref{Beers, T.C., Preston, G.W., \& Shectman, S.A. 1992b}{AJ}{103}{1987
(BPS2)}
\jref{Beers, T.C., Preston, G.W., Shectman, S.A., Doinidis, S.P. \& Griffin,
K.E. 1992a}{AJ}{103}{267}
\jref{Berg, C., Wegner, G., Foltz, C.B., Chaffee, F.H. \& Hewett, P.C. 1992}
{ApJS}{78}{409}
\jref{Bond, H.E. 1980}{ApJS}{44}{517}
\jref{Bowen, D.V. 1991}{MNRAS}{251}{649}
\jref{Burstein, D. \& Heiles, C. 1982}{AJ}{87}{1165}
\jref{Foltz, C.B., Chaffee, F.H, Hewett, P.C., Weymann, R.J. , Anderson, S.F.,
MacAlpine, G.M., 1989}{A.J.}{98}{1959}
\jref{Green, R. F. 1980}{ApJ}{238}{685}
\jref{Greenstein, J.L. 1980}{ApJ}{242}{738}
\jref{Greenstein, J.L. 1984}{ApJ}{276}{602}
\jref{Greenstein, J.L., \& Sargent, A.I. 1974}{ApJS}{28}{157}
\jref{Gunn, J.E. \& Stryker, L.L. 1983}{ApJS}{52}{121}
\jref{Heber, U. 1986}{AAP}{155}{33}
\jref{Herbig, G.H 1992}{Rev. Mexicana AAp}{24}{187}
\jref{Jacoby, G.H., Hunter, D.A., \& Christian, C.A. 1984}{ApJS}{56}{257}
\jref{King, D.J., Birch, I.M., Johnson, C. \& Taylor, K.N.R. 1981}{PASP}{93}
{385}
\ref{Kitchin, C.R. 1984 {\sl Astrophysical Techniques} (Bristol: Adam Hilger)
p.229}
\ref{Miller, J.S. \& Stone, R.P.S. 1987, Lick Observatory Technical
Report 48, {\sl The CCD Cassegrain Spectrograph at the Shane Reflector}}
\jref{Preston, G.W. 1959}{ApJ}{130}{507}
\jref{Sargent, W.L.W., Steidel, C., Boksenberg, A.  1988a}{ApJ}{334}{23}
\jref{Savage, A. et al. 1984}{MNRAS}{207}{393}
\jref{Sion, E.M., Greenstein, J.L., Landstreet, J.D., Liebert, J., Shipman,
H.L.
\& Wegner, G.A. 1983}{ApJ}{269}{253}
\jref{Steidel, C. \& Sargent M. 1992}{ApJS}{80}{1}
\jref{Tytler, D., Boksenberg, A., Sargent, W.L.W.,
Young, P., Kunth, D.  1987} {ApJS}{64}{667}
\jref{Tytler, D., Sandoval, J. \& Fan, X.-M. 1993}{ApJ}{405}{in press}
\jref{Zhan, Y. \& Chen, J-S. 1987}{Act. Ap. Sin.}{7}{99; trans. in
Chin.A.Ap., 11, 191 (ZC1)}
\ref{------. 1987,{Act. Ap. Sin.}{7} 203; trans. in Chin.A.Ap., 11, 299 (ZC2)}
\ref{------. 1989,{Act. Ap. Sin.}{9}  37; trans. in Chin.A.Ap., 13, 139 (ZC3)}
\ref{------. 1989,{Act. Ap. Sin.}{9} 147; trans. in Chin.A.Ap., 13, 321 (ZC4)}
\ref{------. 1989, Chin.A.Ap., 9, 313 (ZC5)}
\vfill\eject
\centerline{FIGURE CAPTIONS}
\bigskip
\noindent Fig. 1.-- Spectra of the QSO candidates. The flux is $f_\nu $, in
units of micro-Jansky. The data have not been smoothed, and individual pixels
are shown. The spectra are grouped by setup (A, then B, then C) and are in
RA order for their setup. Features, both absorption and emission, which are
present in most spectra of a given setup are often caused by erroneous sky
emission line subtraction.

We show enlargements of the blue ends of several of the spectra. When a raised
zero level is required it is shown by the raised horizontal dotted line,
which extends under only the enlargement. The relationship of the flux plotted
in an enlargement to the measured flux is given by the equation under
that enlargement.
\smallskip\noindent Fig. 2.-- The difference between our slit B magnitudes and
ZC's Schmidt plate image size B$\rm _J$ magnitudes. Slit magnitudes from
setup A are shown as open circles, setup B as open squares, and C as filled
dots.
\vfill\eject
%
% -- table 1
%
\baselineskip=13pt
\tabskip=1em
\halign{
# \hfil\tabskip=1em & #\hfil & \hfil#\hfil & \hfil#\hfil & \hfil # \hfil
& \hfil#\hfil &\hfil # \hfil\cr
\multispan{7}\hfil  TABLE 1 \hfil \cr
\multispan{7}\hfil JOURNAL OF OBSERVATIONS \hfil \cr\cr
\noalign{\hrule height .08em \vskip 2pt\hrule height 0.08em \vskip 5pt}
\omit\hfil Object \hfil & \omit\hfil Date \hfil & Integration & Setup
& \multispan{2} \hfil Wavelength Range$^a$ (\AA)\hfil & Ref.$^b$\cr
 &\omit\hfil (U.T.) \hfil & Time (s) & & Blue & Red &  \cr
\noalign{\vskip 5pt \hrule height 0.08em \vskip 5pt}
0000+025A   & 1988--10--04 &  600 & A &\multispan{2}\hfil3843--7022\hfil&
ZC1\cr
0000+027A   & 1990--08--27 & 3600 & B &\multispan{2}\hfil4312--7056\hfil&
ZC1\cr
0002+022    & 1992--10--24 & 2000 & C & 3320--5470 & 5530--8305 & ZC1 \cr
0002+030A   & 1992--10--24 & 2000 & C & 3320--5450 & 5520--8305 & ZC1 \cr
0003+011A   & 1988--10--04 &  600 & A &\multispan{2}\hfil3843--7018\hfil&
ZC1\cr
0003+011B   & 1988--10--04 &  600 & A &\multispan{2}\hfil3843--7018\hfil&
ZC1\cr
0003+012B   & 1992--10--24 & 2000 & C & 3320--5475 & 5530--8305 & ZC1 \cr
0004$-$004  & 1988--10--04 &  600 & A &\multispan{2}\hfil3843--7022\hfil&
ZC2\cr
0004$-$005B & 1990--08--24 & 4000 & B &\multispan{2}\hfil4312--7059\hfil&
ZC2\cr
0004+014    & 1992--10--24 & 2000 & C & 3320--5475 & 5530--8300 & ZC1 \cr
0005+003    & 1992--10--24 & 2000 & C & 3320--5485 & 5530--8310 & ZC1 \cr
0005+030    & 1990--08--24 &  700 & B &\multispan{2}\hfil4312--7059\hfil&
ZC1\cr
0006+020B   & 1990--08--27 & 3000 & B &\multispan{2}\hfil4312--7056\hfil&
ZC1\cr
0006+022A   & 1990--08--27 & 3600 & B &\multispan{2}\hfil4312--7056\hfil&
ZC1\cr
0006+025    & 1990--08--27 & 3600 & B &\multispan{2}\hfil4312--7056\hfil&
ZC1\cr
0008+008    & 1992--10--24 & 2000 & C & 3320--5485 & 5530--8310 & ZC1 \cr
0008+010    & 1992--10--24 & 2000 & C & 3320--5480 & 5530--8310 & ZC1 \cr
0009$-$003  & 1992--10--24 & 2000 & C & 3320--5450 & 5530--8310 & ZC2 \cr
0009+027B   & 1988--10--04 &  600 & A &\multispan{2}\hfil3843--7022\hfil&
ZC1\cr
            & 1990--08--24 &  500 & B &\multispan{2}\hfil4312--7059\hfil&
ZC1\cr
0010$-$002B & 1988--10--04 &  600 & A &\multispan{2}\hfil3843--7006\hfil&
ZC2\cr
0010+008    & 1992--10--24 & 2000 & C & 3320--5485 & 5530--8290 & ZC1 \cr
0010+023    & 1988--10--04 &  600 & A &\multispan{2}\hfil3843--7006\hfil&
ZC1\cr
            & 1990--08--24 & 2300 & B &\multispan{2}\hfil4312--7059\hfil&
ZC1\cr
0011$-$002A & 1988--10--04 &  600 & A &\multispan{2}\hfil3843--7018\hfil&
ZC2\cr
0011$-$002B & 1988--10--04 &  600 & A &\multispan{2}\hfil3843--7018\hfil&
ZC2\cr
0011$-$012A & 1990--08--24 &  600 & B &\multispan{2}\hfil4312--7059\hfil&
ZC2\cr
0011+016    & 1988--10--04 &  600 & A &\multispan{2}\hfil3843--7018\hfil&
ZC1\cr
0012$-$002  & 1990--08--24 &  600 & B &\multispan{2}\hfil4312--7059\hfil&
ZC2\cr
0012+011    & 1990--08--27 & 3000 & B &\multispan{2}\hfil4312--7056\hfil&
ZC1\cr
2348$-$011A & 1992--10--23 &  600 & C & 3320--5485 & 5530--8310 & ZC4 \cr
2349$-$012  & 1992--10--23 & 1000 & C & 3320--5485 & 5530--8310 & ZC4 \cr
2350$-$019A & 1992--10--23 & 1000 & C & 3320--5480 & 5540--8275 & ZC4 \cr
2356+007    & 1992--10--23 & 1000 & C & 3320--5485 & 5530--8310 & ZC3 \cr
2357$-$005A & 1992--10--23 & 2000 & C & 3320--5485 & 5530--8275 & ZC4 \cr
2357+009    & 1992--10--23 & 2000 & C & 3320--5485 & 5520--8300 & ZC3  \cr
2358$-$011  & 1992--10--23 & 2000 & C & 3320--5480 & 5530--8310 & ZC4 \cr
2358+004    & 1992--10--23 & 2000 & C & 3320--5450 & 5530--8310 & ZC3 \cr
2358+029A   & 1992--10--25 & 2000 & C & 3320--5480 & 5530--8295 & ZC3 \cr
2359+026    & 1992--10--25 & 2000 & C & 3320--5485 & 5530--8270 & ZC3 \cr
\noalign{\vskip 5pt \hrule height .1em \vskip 1em}
}
\noindent
$^a$ UV Schmidt has only one wavelength setting, whereas Kast is a double
spectrograph.

\noindent
$^b$ Reference for coordinates and finding charts.
\vfill\eject
%
% -- table 2
%
\baselineskip=14pt
\def\Ha{H$\alpha$}
\def\Hb{H$\beta$}
\def\Hg{H$\gamma$}
\def\Lya{Ly$\alpha$}
\def\spa{\omit\hfil ... \hfil}
% begin table
\tabskip=1em
\halign{
# \hfil\tabskip=0.8em & \hfil # \hfil & \hfil#\hfil & \hfil#\hfil
& # \hfil & # \hfil \cr
\multispan{6}\hfil  TABLE 2 \hfil \cr
\multispan{6}\hfil SUMMARY OF SPECTRA \hfil \cr\cr
\noalign{\hrule height .08em \vskip 2pt\hrule height 0.08em \vskip 5pt}
\omit\hfil Object \hfil & Prism z$_{\rm em}^{~a}$ & Q~$^a$ & Type~$^b$
&\omit\hfil z$_{\rm em}$ \hfil  & Spectral Lines \cr
\noalign{\vskip 5pt \hrule height 0.08em \vskip 5pt}
0000+025A   & 3.15  & 5 & QSO      & 1.6843 & C~IV, C~III] emission \cr
0000+027A   & 1.71  & 5 & QSO      & 2.3840 & C~IV, C~III] emission \cr
0002+022    & 3.17  & 5 & star:F(F7) & \spa & Ca~II H \& K, \Ha, \Hb, \Hg\ etc.
absorption \cr
0002+030A   & 3.15  & 5 & star:G(F8) & \spa & Ca~II H \& K, \Ha, G-band
absorption \cr
0003+011A   & \spa  & 4 & star:M5(M2)& \spa & Ti~O, Na~I absorption \cr
0003+011B   & 3.17  & 5 & star:F(F3) & \spa & \Hb\ absorption, weak or absent
Ca~II, \Ha, \Hg \cr
0003+012B   & 3.12  & 6 & star:G(G0) & \spa & Ca~II H \& K, \Ha, \Hb, Mg~I,
G-band absorption\cr
0004$-$004  & 3.12  & 5 & star:K(K0) & \spa & Ca~II H \& K, G-band absorption
\cr
0004$-$005B & 1.75  & 5 & QSO      & 1.7195 &  He~II, O~III] and C~III]
emission \cr
0004+014    & 3.14  & 5 & star:G/K(K1)&\spa & Ca~II H \& K, \Ha, Mg~I, G-band
absorption \cr
0005+003    & 3.00  & 5 & galaxy   & 0.0932 & Ca~II H \& K redshifted \cr
0005+030    & 2.26  & 5 & QSO      & 1.0948 & Mg~II, Fe~II emission \cr
0006+020B   & 1.70  & 5 & QSO      & 2.3483 & Si~IV+O~VI], C~IV and C~III]
emission \cr
0006+022A   & 2.22  & 6 & QSO      & 1.5152 & C~III] emission \cr
0006+025    & 2.14  & 5 & QSO      & 2.0909 & C~IV, C~III] emission \cr
0008+008    & 3.15  & 5 & QSO      & 3.0837 & Ly$\alpha$, C~IV, C~III] emission
\cr
0008+010    & 3.12  & 5 & star:G(F8) & \spa & Ca~II H \& K, \Ha, \Hb, \Hg\
absorption,
no G-band \cr
0009$-$003  & 3.02  & 5 & star:F(F6) & \spa & Ca~II H \& K, \Ha, \Hb, \Hg\ etc.
absorption \cr
0009+027B   & 3.17  & 5 & star:G(F5--G0)&\spa & Ca~II H \& K, \Ha, \Hb\
absorption \cr
0010$-$002B & 2.11  & 5 & QSO      & 2.1453 & C~IV \& C~III] emission \cr
0010+008    & 3.11  & 5 & QSO      & 3.076  & \Lya, N~V, Si~IV+O~VI], C~III]
and C~IV emission \cr
0010+023    & 2.27  & 5 & galaxy   & 0.0879 & \Hb, \Hg, [O~II], O~III] etc.
emission \cr
0011$-$002A & 3.10  & 5 & galaxy   & 0.1158 & [O~II] \& \Hb\ emission \cr
0011$-$002B & \spa  & 4 & star:K(K4) & \spa & Ca~II H \& K, \Ha, \Hb, Mg~I,
G-band  absorption \cr
0011$-$012A & 2.23  & 5 & star:sdB(A6)&\spa & strong, broad \Ha, \Hb, \Hg\
absorption, no He~I~$\lambda$4471 \cr
0011+016    & 3.22  & 5 & star:G(F8) & \spa & Ca~II H \& K, \Ha, \Hb, \Hg\
absorption \cr
0012$-$002  & 2.28  & 5 & QSO      & 1.557? & possible C~III] emission \cr
0012+011    & 1.67  & 5 & star:K(K4) & \spa & possible Na~I absorption, low S/N
 \cr
2348$-$011A & 3.07  & 5 & star:G(F5) & \spa & Ca~II H \& K absorption \cr
2349$-$012  & 3.12  & 5 & star:G(G8) & \spa & Ca~II H \& K, \Ha, \Hb\
absorption \cr
2350$-$019A & 3.10  & 5 & WD:DA4(A3) & \spa & Balmer absorption up to H5 \cr
2356+007    & 3.17  & 5 & star:M3(M3)& \spa & Ti~O, Ca~II H \& K absorption \cr
2357$-$005A & 2.83  & 5 & galaxy   & 0.106  & \Ha, [O~II] emission \cr
2357+009    & 3.05  & 5 & star:G(G2) & \spa & Ca~II H \& K, \Ha, \Hb\
absorption \cr
2358$-$011  & 3.22  & 5 & star:G(F7) & \spa & Ca~II H \& K, \Ha, \Hb\
absorption \cr
2358+004    & 2.94  & 5 & star:G(G3) & \spa & Ca~II H \& K, \Ha\ absorption \cr
2358+029A   & 3.13  & 5 & star:M3(M5)& \spa & Ti~O, Na~I absorption \cr
2359+026    & 3.14  & 5 & ?:?(A7)    & \spa & low S/N flat spectrum,
featureless \cr
\noalign{\vskip 5pt \hrule height .1em \vskip 1em}
}
\noindent
$^a$ Objective prism redshift and the reliability index Q are from the ZC
references
listed in Table 1.

\noindent
$^b$ The first stellar type is from the spectral features and the second, in
parenthesis,
is from (B$-$V)$_{\rm 0}$~.
\vfill\eject
%
% -- table 3
%
\baselineskip=12pt
\def\spa{\omit\hfil ... \hfil}
\tabskip=1em
\halign{
# \hfil\tabskip=6pt &\hfil # \hfil & \hfil # \hfil & \hfil#\hfil & \hfil#\hfil
& \hfil # \hfil & \hfil # \hfil & \hfil#  & \hfil # \hfil & \hfil#
& \hfil#\hfil \cr
\multispan{11}\hfil  TABLE 3 \hfil \cr
\multispan{11}\hfil MAGNITUDES AND SPECTRAL INDICES \hfil \cr\cr
\noalign{\hrule height .08em \vskip 2pt\hrule height 0.08em \vskip 5pt}
\omit\hfil Object \hfil & B$_J^{~a}$ & U~$^b$  & B~$^b$ & V~$^b$
& (B$-$V)$_0~^b$ & (B$-$V)~$^c$ &\omit\hfil $\Delta~^d$ \hfil & HP~$^e$
& \omit\hfil KP$^{\prime~e}$ \hfil & GP~$^e$ \cr
\noalign{\vskip 5pt \hrule height 0.08em \vskip 5pt}
0000+025A$^g$   & 17.5 & \spa  & 18.88 & 18.71 & 0.15 & \spa & \spa  & \spa &
\spa & \spa  \cr
0000+027A       & 18.0 & \spa  & 21.03 & 20.14 & 0.87 & \spa & \spa  & \spa &
\spa & \spa  \cr
0002+022        & 18.0 & 18.28 & 18.64 & 18.15 & 0.48 & 0.53 &$-$0.05& 1.92 &
5.23 & 0.69  \cr
0002+030A       & 18.5 & 19.85 & 20.26 & 19.70 & 0.54 & \spa & \spa  & \spa &
\spa & \spa  \cr
0003+011A$^g$   & 18.0 & \spa  & 19.75 & 18.22 & 1.51 & \spa & \spa  & \spa &
\spa & \spa  \cr
0003+011B$^g$   & 18.0 & \spa  & 18.77 & 18.44 & 0.42 & \spa & \spa  & \spa &
\spa & \spa  \cr
0003+012B       & 18.0 & 18.43 & 18.69 & 18.12 & 0.55 & 0.62 &$-$0.07& 1.44 &
7.91 & 3.03  \cr
0004$-$004$^g$  & 17.5 & \spa  & 18.58 & 17.79 & 0.78 & \spa & \spa  & \spa &
\spa & \spa  \cr
0004$-$005B     & 18.0 & \spa  & 18.70 & 18.30 & 0.39 & \spa & \spa  & \spa &
\spa & \spa  \cr
0004+014        & 18.5 & 19.68 & 19.31 & 18.39 & 0.90 & 0.85 &  0.05 & 0.39 &
9.43 & 5.35  \cr
0005+003        & 17.5 & 19.83 & 19.47 & 18.23 & 1.22 & \spa & \spa  & \spa &
\spa & \spa  \cr
0005+030        & 17.0 & \spa  & 16.72 & 16.10 & 0.60 & \spa & \spa  & \spa &
\spa & \spa  \cr
0006+020B       & 17.5 & \spa  & 18.44 & 17.75 & 0.68 & \spa & \spa  & \spa &
\spa & \spa  \cr
0006+022A       & 18.0 & \spa  & 19.49 & 19.11 & 0.37 & \spa & \spa  & \spa &
\spa & \spa  \cr
0006+025        & 17.5 & \spa  & 19.19 & 18.75 & 0.43 & \spa & \spa  & \spa &
\spa & \spa  \cr
0008+008        & 19.5 & 21.53 & 20.05 & 18.87 & 1.16 & \spa & \spa  & \spa &
\spa & \spa  \cr
0008+010        & 19.0 & 19.02 & 19.32 & 18.77 & 0.53 & 0.53 &  0.00 & 1.92 &
2.94 & 1.45  \cr
0009$-$003      & 18.0 & 18.41 & 18.77 & 18.29 & 0.47 & 0.45 &  0.02 & 2.55 &
6.46 & 1.05  \cr
0009+027B$^f$   & 16.0 & \spa & 17.55 & 16.95 & 0.58 & \spa & \spa  & \spa &
\spa & \spa  \cr
0009+027B$^g$   & 16.0 & \spa & 18.39 & 17.93 & 0.44 & \spa & \spa  & \spa &
\spa & \spa  \cr
0010$-$002B$^g$ & 18.0 & \spa  & 20.07 & 19.43 & 0.63 & \spa & \spa  & \spa &
\spa & \spa  \cr
0010+008        & 19.5 & 21.84 & 20.42 & 19.67 & 0.74 & \spa & \spa  & \spa &
\spa & \spa  \cr
0010+023$^f$    & 15.5 & \spa  & 17.85 & 17.25 & 0.58 & \spa & \spa  & \spa &
\spa & \spa  \cr
0010+023$^g$    & 15.5 & \spa  & 18.53 & 17.95 & 0.56 & \spa & \spa  & \spa &
\spa & \spa  \cr
0011$-$002A$^g$ & 17.5 & \spa  & 19.37 & 18.19 & 1.17 & \spa & \spa  & \spa &
\spa & \spa  \cr
0011$-$002B$^g$ & 17.5 & \spa  & 18.41 & 17.36 & 1.04 & \spa & \spa  & \spa &
\spa & \spa  \cr
0011$-$012A     & 17.0 & \spa  & 17.90 & 17.69 & 0.18 & \spa & \spa  & \spa &
\spa & \spa  \cr
0011+016$^g$    & 17.5 & \spa  & 18.28 & 17.73 & 0.54 & \spa & \spa  & \spa &
\spa & \spa  \cr
0012$-$002      & 18.0 & \spa  & 17.60 & 17.19 & 0.40 & \spa & \spa  & \spa &
\spa & \spa  \cr
0012+011        & 18.0 & \spa  & 19.19 & 18.10 & 1.08 & \spa & \spa  & \spa &
\spa & \spa  \cr
2348$-$011A     & 18.5 & 19.09 & 19.33 & 18.86 & 0.45 & 0.47 &$-$0.02& 2.36 &
3.74 & 2.51  \cr
2349$-$012      & 16.5 & 18.62 & 18.59 & 17.85 & 0.73 & 0.71 &  0.02 & 0.85 &
8.05 & 4.07  \cr
2350$-$019A     & 18.5 & 17.89 & 18.80 & 18.70 & 0.09 & \spa & \spa  & \spa &
\spa & \spa  \cr
2356+007        & 18.5 & 20.17 & 19.29 & 17.73 & 1.54 & \spa & \spa  & \spa &
\spa & \spa  \cr
2357$-$005A     & 18.0 & 20.33 & 20.83 & 20.07 & 0.74 & \spa & \spa  & \spa &
\spa & \spa  \cr
2357+009        & 17.0 & 20.19 & 20.40 & 19.71 & 0.66 & 0.61 &  0.05 & 1.46 &
5.68 & 5.03  \cr
2358$-$011      & 17.5 & 19.25 & 19.35 & 18.81 & 0.52 & 0.46 &  0.06 & 2.51 &
4.07 & 2.26  \cr
2358+004        & 17.0 & 19.51 & 19.57 & 18.89 & 0.66 & 0.64 &  0.02 & 1.33 &
11.88& 5.10  \cr
2358+029A       & 17.5 & 21.34 & 19.87 & 18.16 & 1.69 & \spa & \spa  & \spa &
\spa & \spa  \cr
2359+026        & 17.0 & 19.33 & 19.99 & 19.61 & 0.36 & \spa & \spa  & \spa &
\spa & \spa  \cr
\noalign{\vskip 5pt \hrule height .1em \vskip 1em}
}
\noindent
$^a$ From the ZC reference listed in Table 1.

\noindent
$^b$ Highly uncertain because they are derived from narrow slit spectra,
discussed in \S3.2.

\noindent
$^c$ Estimated from the Balmer line index HP using eqn. (3).

\noindent
$^d$ Slit (B$-$V) minus HP index (B$-$V).

\noindent
$^e$ The H$\delta$, Ca~II~K and G-band equivalent width (\AA), defined in
\S4.1.

\noindent
$^f$ Magnitudes were measured from Setup B spectrum.

\noindent
$^g$ Magnitudes were  measured from Setup A spectrum, and likely to be
systematically too faint.
\vfill\eject
%
% -- table-4
%

\baselineskip=20pt
\def\spa{\omit\hfil ... \hfil}
% begin table
\tabskip=1em
$$\vbox{
\halign{
# \hfil\tabskip=1em & # \hfil & \hfil #  & #\hfil & \hfil#  \cr
\multispan{5}\hfil  TABLE 4 \hfil \cr
\multispan{5}\hfil STELLAR ABUNDANCES AND DISTANCES \hfil \cr\cr
\noalign{\hrule height .08em \vskip 2pt\hrule height 0.08em \vskip 5pt}
\omit\hfil Object \hfil & \multispan{2} \hfil Giant \hfil
& \multispan{2} \hfil Dwarf \hfil \cr
 &\multispan{2}\hrulefill &\multispan{2} \hrulefill \cr
 & \omit\hfil [Fe/H] \hfil & \omit\hfil d(kpc) \hfil
   & \omit\hfil [Fe/H] \hfil & \omit\hfil d(kpc) \hfil  \cr
\noalign{\vskip 5pt \hrule height 0.08em \vskip 5pt}
0002+022    & $-$1.7$\pm$0.2 &  9.0~$^a$ & $-$1.7$\pm$0.2 &  3.7 \cr
0003+012B   & $-$1.3$\pm$0.1 &  8.5~$^a$ & $-$1.4$\pm$0.2 &  2.9 \cr
0004+014    & $-$1.9$\pm$0.1 & 62.2~~    & $-$2.1$\pm$0.1 &  1.4 \cr
0008+010    & $-$2.5         & 14.4~$^a$ & $-$2.6         &  4.0 \cr
0009$-$003  & $-$1.0$\pm$0.2 &  6.7~$^a$ & $-$1.0$\pm$0.2 &  6.6 \cr
2348$-$011A & $-$1.8$\pm$0.2 & 12.1~$^a$ & $-$2.0$\pm$0.1 &  5.8 \cr
2349$-$012  & $-$1.8$\pm$0.1 & 23.3~~    & $-$2.0$\pm$0.1 &  1.6 \cr
2357+009    & $-$2.2$\pm$0.1 & 42.7~~    & $-$2.2$\pm$0.1 &  4.6 \cr
2358$-$011  & $-$1.9$\pm$0.3 & 12.8~$^a$ & $-$2.0$\pm$0.1 &  4.9 \cr
2358+004    & $-$0.5:        &  \spa\ \  & $-$0.5:        &  ...  \cr
\noalign{\vskip 5pt \hrule height .1em \vskip 1em}
\noalign{\hbox{$^a$ Subgiants, near the main sequence turnoff.}}
}
}$$
\vfill\eject
%
% -- table 5
%

\baselineskip=13pt
\def\Ha{H$\alpha$}
\def\Hb{H$\beta$}
\def\Hg{H$\gamma$}
\def\Lya{Ly$\alpha$}
% begin table
$$\vbox{
\tabskip=1em
\halign{
# \hfil\tabskip=1em & # \hfil & \hfil#\hfil & \hfil# &\hfil # \hfil \cr
\multispan{5}\hfil  TABLE 5 \hfil \cr
\multispan{5}\hfil EMISSION LINES \hfil \cr\cr
\noalign{\hrule height .08em \vskip 2pt\hrule height 0.08em \vskip 5pt}
\omit\hfil Ion \hfil & \omit\hfil $\lambda_{\rm lab}~^a$ \hfil
& $\lambda_{\rm obs}~^a$ & W$_{\rm obs}$ & z$_{\rm em}$ \cr
\noalign{\vskip 5pt \hrule height 0.08em \vskip 5pt}
\multispan{5} \hfil 0000+025A \hfil \cr
\noalign{\vskip 5pt \hrule height 0.08em \vskip 5pt}
C~IV   & 1549.06 & 4158.4 & 70 & 1.6845 \cr
C~III] & 1908.73 & 5123.2 & 56 & 1.6841 \cr
\multispan{5} \hfil $<z_{em}>$=1.6843 \hfil \cr
\noalign{\vskip 5pt \hrule height 0.08em \vskip 5pt}
\multispan{5} \hfil 0000+027A \hfil \cr
\noalign{\vskip 5pt \hrule height 0.08em \vskip 5pt}
C~IV   & 1549.06 & 5237.9 & 98 & 2.3835 \cr
C~III] & 1908.73 & 6460.3 & 77 & 2.3846 \cr
\multispan{5} \hfil $<z_{em}>$=2.384 \hfil \cr
\noalign{\vskip 5pt \hrule height 0.08em \vskip 5pt}
\multispan{5} \hfil 0004$-$005B \hfil \cr
\noalign{\vskip 5pt \hrule height 0.08em \vskip 5pt}
He~II  & 1640.43 & 4462.0 &  8 & 1.7200 \cr
O~III] & 1663.99 & 4537.3 &  8 & 1.7268 \cr
C~III] & 1908.73 & 5187.9 & 42 & 1.7180 \cr
\multispan{5} \hfil $<z_{em}>$=1.7195 \hfil \cr
\noalign{\vskip 5pt \hrule height 0.08em \vskip 5pt}
\multispan{5} \hfil 0005+003 (absorption lines only) \hfil \cr
\noalign{\vskip 5pt \hrule height 0.08em \vskip 5pt}
Ca~II  & 3934.78 & 4301.6 & 14.6 & 0.0932 \cr
Ca~II  & 3969.59 & 4339.5 & 14.6 & 0.0932 \cr
G-band & 4301.2: & 4706.5 &  6.2 & 0.0942 \cr
\Hb    & 4862.68 & 5315.6 &  5.3 & 0.0931 \cr
Na~I~D & 5894.56 & 6442.2 &  2.6 & 0.0929 \cr
\multispan{5} \hfil $<z_{\rm abs}>$=0.0932 \hfil \cr
\noalign{\vskip 5pt \hrule height 0.08em \vskip 5pt}
\multispan{5} \hfil 0005+030~$^b$ \hfil \cr
\noalign{\vskip 5pt \hrule height 0.08em \vskip 5pt}
Mg~II & 2798.74 & 5863.0 & 44 & 1.0948 \cr
\noalign{\vskip 5pt \hrule height 0.08em \vskip 5pt}
\multispan{5} \hfil 0006+020B \hfil \cr
\noalign{\vskip 5pt \hrule height 0.08em \vskip 5pt}
Si~IV+O~IV] & 1398.62 & 4693.0 & 22 & 2.3555 \cr
C~IV   & 1549.06 & 5186.7 & 38 & 2.3483 \cr
C~III] & 1908.73 & 6391.0 & 50 & 2.3483 \cr
\multispan{5} \hfil $<z_{em}>$=2.3483 \hfil \cr
\noalign{\vskip 5pt \hrule height 0.08em \vskip 5pt}
\multispan{5} \hfil 0006+022A~$^b$ \hfil \cr
\noalign{\vskip 5pt \hrule height 0.08em \vskip 5pt}
C~III]  & 1908.73 & 4800.9 & 46 & 1.5152 \cr
\noalign{\vskip 5pt \hrule height 0.08em \vskip 5pt}
}}$$
\vfill\eject

$$\vbox{
\halign{
# \hfil\tabskip=1em & # \hfil & \hfil#\hfil & \hfil# &\hfil # \hfil \cr
\multispan{5}\hfil  TABLE 5 -- {\it continued} \hfil \cr\cr
\noalign{\hrule height .08em \vskip 2pt\hrule height 0.08em \vskip 5pt}
\omit\hfil Ion \hfil & \omit\hfil $\lambda_{\rm lab}~^a$ \hfil
& $\lambda_{\rm obs}~^a$ & W$_{\rm obs}$ & z$_{\rm em}$ \cr
\noalign{\vskip 5pt \hrule height 0.08em \vskip 5pt}
\multispan{5} \hfil 0006+025 \hfil \cr
\noalign{\vskip 5pt \hrule height 0.08em \vskip 5pt}
C~IV   & 1549.06 & 4787.5 & 102 & 2.0906 \cr
C~III] & 1908.73 & 5900.4 &  57 & 2.0913 \cr
\multispan{5} \hfil $<z_{em}>$=2.0909 \hfil \cr
\noalign{\vskip 5pt \hrule height 0.08em \vskip 5pt}
\multispan{5} \hfil 0008+008 \hfil \cr
\noalign{\vskip 5pt \hrule height 0.08em \vskip 5pt}
Ly$\alpha$ & 1215.67 & 4977.4 & 285 & 3.0944~$^b$ \cr
C~IV       & 1549.06 & 6324.8 & 163 & 3.0830 \cr
O~III]     & 1643.99 & 6794.7 &  25 & 3.0843 \cr
C~III]     & 1908.73 & 7796.7 & 100 & 3.0848 \cr
\multispan{5} \hfil $<z_{em}>$=3.0837 \hfil \cr
\noalign{\vskip 5pt \hrule height 0.08em \vskip 5pt}
\multispan{5} \hfil 0010$-$002B \hfil \cr
\noalign{\vskip 5pt \hrule height 0.08em \vskip 5pt}
C~IV   & 1549.06 & 4873.1 & 40 & 2.1458 \cr
C~III] & 1908.73 & 6005.6 & 87 & 2.1464 \cr
\multispan{5} \hfil $<z_{em}>$=2.1462 \hfil \cr
\noalign{\vskip 5pt \hrule height 0.08em \vskip 5pt}
\multispan{5} \hfil 0010+008 \hfil \cr
\noalign{\vskip 5pt \hrule height 0.08em \vskip 5pt}
Ly$\beta$  & 1025.72 & 4190.2 & 102 & 3.0851 \cr
Ly$\alpha$ & 1215.67 & 4963.7 & 148 & 3.0831 \cr
N~V        & 1240.13 & 5048.9 &  40 & 3.0712 \cr
Si~IV+O~IV]& 1398.64 & 5688.3 &  45 & 3.0670 \cr
C~IV       & 1549.06 & 6303.6 &  94 & 3.0693 \cr
C~III]     & 1908.73 & 7786.6 &  65 & 3.0795 \cr
\multispan{5} \hfil $<z_{em}>$=3.076 \hfil \cr
\noalign{\vskip 5pt \hrule height 0.08em \vskip 5pt}
\multispan{5} \hfil 0010+023 \hfil \cr
\noalign{\vskip 5pt \hrule height 0.08em \vskip 5pt}
\multispan{5} \hfil Setup A \hfil \cr
\noalign{\vskip 5pt \hrule height 0.08em \vskip 5pt}
[O~II]      & 3728.06 & 4065.6 & 36.8 & 0.0905 \cr
[Ne~III]    & 3870.10 & 4209.7 &  8.5 & 0.0877 \cr
\Hg         & 4341.68 & 4726.8 &  4.9 & 0.0887 \cr
\Hb         & 4862.68 & 5291.3 & 17.2 & 0.0881 \cr
[O~III]     & 4960.28 & 5398.5 &  8.4 & 0.0883 \cr
[O~III]     & 5008.20 & 5449.1 & 18.9 & 0.0880 \cr
He~I        & 5877.63 & 6393.3 &  5.0 & 0.0877 \cr
\noalign{\vskip 5pt \hrule height 0.08em \vskip 5pt}
\multispan{5} \hfil $<z_{em}>$=0.0889 \hfil \cr
\noalign{\vskip 5pt \hrule height 0.08em \vskip 5pt}
}}$$
\vfill\eject
$$\vbox{
\halign{
# \hfil\tabskip=1em & # \hfil & \hfil#\hfil & \hfil# &\hfil # \hfil \cr
\multispan{5}\hfil  TABLE 5 -- {\it continued} \hfil \cr\cr
\noalign{\hrule height .08em \vskip 2pt\hrule height 0.08em \vskip 5pt}
\omit\hfil Ion \hfil & \omit\hfil $\lambda_{\rm lab}~^a$ \hfil
& $\lambda_{\rm obs}~^a$ & W$_{\rm obs}$ & z$_{\rm em}$ \cr
\noalign{\vskip 5pt \hrule height 0.08em \vskip 5pt}
\multispan{5} \hfil Setup B (better spectrum) \hfil \cr
\noalign{\vskip 5pt \hrule height 0.08em \vskip 5pt}
\Hg         & 4341.68 & 4721.6 &  1.3 & 0.0875 \cr
\Hb         & 4862.68 & 5290.1 & 15.6 & 0.0879 \cr
[O~III]     & 4960.28 & 5396.3 &  5.7 & 0.0879 \cr
[O~III]     & 5008.20 & 5448.4 & 16.6 & 0.0879 \cr
He~I        & 5877.63 & 6393.2 &  2.6 & 0.0877 \cr
[O~I]+[S~III] & 6301.74  & 6855.7 &  3.0 & 0.0879 \cr
\multispan{5} \hfil $<z_{em}>$=0.0879 \hfil \cr
\noalign{\vskip 5pt \hrule height 0.08em \vskip 5pt}
\multispan{5} \hfil 0011$-$002A~$^b$ \hfil \cr
\noalign{\vskip 5pt \hrule height 0.08em \vskip 5pt}
[O~II]   & 3728.06    & 4159.9  & 13.8 & 0.1158 \cr
\Hb      & 4862.68 & 5424.6  & 4.7 & 0.1156 \cr
\multispan{5} \hfil $<z_{em}>$=0.1158 \hfil \cr
\noalign{\vskip 5pt \hrule height 0.08em \vskip 5pt}
\multispan{5} \hfil 0012$-$002~$^b$ \hfil \cr
\noalign{\vskip 5pt \hrule height 0.08em \vskip 5pt}
C~III]        & 1908.73 & 4881.7 & 26   & 1.5576 \cr
\noalign{\vskip 5pt \hrule height 0.08em \vskip 5pt}
\multispan{5} \hfil 2357$-$005A \hfil \cr
\noalign{\vskip 5pt \hrule height 0.08em \vskip 5pt}
[O~II]        & 3728.06 & 4122.6 & 23.4 & 0.1058 \cr
\Ha           & 6564.56 & 7257.9 & 37   & 0.1056 \cr
[N~II]        & 6584.82 & 7281.9 &  9.8 & 0.1059 \cr
[S~II]        & 6718.85 & 7428.5 & 17   & 0.1056 \cr
[S~II]        & 6735.86 & 7443.1 & 19   & 0.1050 \cr
\multispan{5} \hfil $<z_{em}>$=0.1056 \hfil \cr
\noalign{\vskip 5pt \hrule height .1em \vskip 1em}
\noalign{\hbox{$^a$ These are vaccum wavelength.}\vskip 5pt}
\noalign{\hbox{$^b$ See notes on individual objects.}}
}
}$$
\vfill\eject
%
% -- table 6
%
\baselineskip=15pt
\def\spa{\omit\hfil ... \hfil}
% begin table
$$\vbox{
\tabskip=1em
\halign{
# \hfil\tabskip=1em & \hfil# & \hfil#\hfil & \hfil # & #\hfil &\hfil #\hfil\cr
\multispan{6}\hfil  TABLE 6 \hfil \cr
\multispan{6}\hfil MAJOR ABSORPTION LINES IN THE QSO SPECTRA \hfil \cr\cr
\noalign{\hrule height .08em \vskip 2pt\hrule height 0.08em \vskip 5pt}
\omit\hfil QSO \hfil & \omit\hfil No. \hfil & $\lambda_{\rm obs}~^a$ &
\omit\hfil W$_{\rm obs}$ & Identification  & z$_{\rm abs}$ \cr
\noalign{\vskip 5pt \hrule height 0.08em \vskip 5pt}
0000+025A   &  1 & 6382.0 & 11.2 & \spa                      & \spa  \cr
0000+027A   &  1 & 5257.3 & 10.1 & C~IV~($\lambda$1548)~$^b$ & 2.394 \cr
0004$-$005B &  1 & 5854.3 &  4.5 &  \spa                     & \spa  \cr
            &  2 & 5959.2 &  5.2 &  \spa                     & \spa  \cr
0006+020B   &  1 & 4700.6 &  7.1 & C~IV~($\lambda$1548)~$^{b, c}$ & 2.034 \cr
            &    &        &      & Si~IV~($\lambda$1393)~$^c$& 2.373 \cr
            &  2 & 4735.5 &  0.6 & Si~IV~($\lambda$1402)~$^c$& 2.376 \cr
            &  3 & 5073.8 &  1.7 & Al~II~($\lambda$1670)~$^c$& 3.037 \cr
            &    &        &      & Fe~II~($\lambda$2382)~$^c$& 1.129 \cr
            &  4 & 5225.9 &  2.7 & C~IV~($\lambda$1548)~$^{b, c}$ & 2.374 \cr
            &  5 & 5578.3 &  2.0 &  \spa                     & \spa  \cr
            &  6 & 5960.7 &  1.2 & Mg~II~($\lambda$2796)     & 1.131 \cr
            &  7 & 5975.5 &  0.7 & Mg~II~($\lambda$2803)     & 1.131 \cr
0006+025    &  1 & 6438.8 &  2.7 &  \spa                     & \spa  \cr
0008+008    &  1 & 4883.5 & 96.3 & H~I~($\lambda$1215)       & 3.017 \cr
            &  2 & 4962.0 & 12.3 & H~I~($\lambda$1215)       & 3.082 \cr
            &  3 & 5613.7 &  3.6 & C~IV~($\lambda$1548)      & 2.626 \cr
            &    &        &      & Si~IV($\lambda$1393)      & 3.028 \cr
            &  4 & 5622.2 &  1.2 & C~IV~($\lambda$1550)      & 2.625 \cr
            &  5 & 5650.3 &  3.2 & C~IV~($\lambda$1548)      & 2.650 \cr
            &    &        &      & Si~IV($\lambda$1402)      & 3.028 \cr
            &  6 & 5659.5 &  1.6 & C~IV~($\lambda$1550)      & 2.650 \cr
            &  7 & 6030.0 &  1.8 & C~IV~($\lambda$1548)      & 2.895 \cr
            &  8 & 6039.3 &  0.7 & C~IV~($\lambda$1550)      & 2.894 \cr
            &  9 & 6134.1 &  2.1 & Mg~II~($\lambda$2796)     & 1.194 \cr
            & 10 & 6150.4 &  2.3 & Mg~II~($\lambda$2803)     & 1.194 \cr
            & 11 & 6238.0 &  7.0 & C~IV~($\lambda$1548)      & 3.029 \cr
            & 12 & 6248.6 &  4.1 & C~IV~($\lambda$1550)      & 3.029 \cr
            & 13 & 6314.1 &  6.8 & C~IV~($\lambda$1548)      & 3.078 \cr
            & 14 & 6325.7 &  4.3 & C~IV~($\lambda$1550)      & 3.079 \cr
            & 15 & 6730.6 &  3.9 & Al~II~($\lambda$1670)     & 3.028 \cr
            & 16 & 7471.9 &  1.4 & Al~III~($\lambda$1854)    & 3.028 \cr
            & 17 & 7503.7 &  3.1 & Al~III~($\lambda$1862)    & 3.028 \cr
\noalign{\vskip 5pt \hrule height .1em \vskip 5pt}
}}$$ \vfill\eject
$$\vbox{
\tabskip=1em
\halign{
# \hfil\tabskip=1em & \hfil# & \hfil#\hfil & \hfil # & #\hfil &\hfil #\hfil\cr
\multispan{6}\hfil  TABLE 6 ---{\it continued}\hfil \cr\cr
\noalign{\hrule height .08em \vskip 2pt\hrule height 0.08em \vskip 5pt}
\omit\hfil QSO \hfil & \omit\hfil No. \hfil & $\lambda_{\rm obs}~^a$ &
\omit\hfil W$_{\rm obs}$ & Identification  & z$_{\rm abs}$ \cr
\noalign{\vskip 5pt \hrule height 0.08em \vskip 5pt}
0010$-$002B &  0 & 3669~$^d$ &...& H~I~($\lambda$1216)       & 2.017 \cr
	      &  1 & 4691.3 &  7.0 & C~IV~($\lambda$1548)      & 2.030 \cr
            &  2 & 4699.0 &  6.3 & C~IV~($\lambda$1550)      & 2.030 \cr
            &  3 & 5075.4 &  9.0 & \spa                      & \spa  \cr
	      &  4 & 6286.4 &  3.9 & Mg~II~($\lambda$2976)     & 1.248 \cr
            &  5 & 6300.6 &  5.2 & Mg~II~($\lambda$2803)     & 1.247 \cr
\noalign{\vskip 5pt \hrule height .1em \vskip 5pt}
\noalign{\hbox{$^a$ Vaccum wavelength.}}
\noalign{\hbox{$^b$ Blended with C~IV~$\lambda$1550.}}
\noalign{\hbox{$^c$ See \S6.}}
\noalign{\hbox{$^d$ Seen by ZC2.}}
}
}$$
\bye